\documentclass[aps,pre,a4paper,amsmath,twocolumn,amssymb]{revtex4}
\usepackage{epsfig}
\usepackage{amssymb}
\usepackage{graphicx}
\usepackage{dcolumn}
\usepackage{bm}

\begin{document}
\title{Biaxial nematic and smectic phases of parallel particles 
with different cross sections}
\author{Yuri Mart\'{\i}nez-Rat\'on}
\email{yuri@math.uc3m.es}
\affiliation{Grupo Interdisciplinar de Sistemas Complejos (GISC), 
Departamento de Matem\'{a}ticas,
Escuela Polit\'{e}cnica Superior, Universidad Carlos III de Madrid,
Avenida de la Universidad 30, E--28911, Legan\'{e}s, Madrid, Spain
}
\author{Szabolcs Varga}
\email{vargasz@almos.vein.hu}
\affiliation{Department of Physics, University of Veszpr\'em, 
H-8201 Veszpr\'em, P. O. Box 158, Hungary}
\author{Enrique Velasco}
\email{enrique.velasco@uam.es}
\affiliation{Departamento de F\'{\i}sica Te\'orica de la Materia Condensada
and Instituto de Ciencia de Materiales Nicol\'as Cabrera,
Universidad Aut\'onoma de Madrid, E-28049 Madrid, Spain.}

\date{\today}

\begin{abstract}
We have calculated the phase diagrams of one--component 
fluids made of five types of biaxial particles differing in their cross 
sections. The orientation of the
principal particle axis is fixed in space, while the second axis is allowed 
to freely rotate. We have constructed a free-energy density functional based on 
fundamental--measure theory to study the relative stability of  
nematic and smectic phases with uniaxial, biaxial and tetratic symmetries.
Minimization of the density functional allows us to study the phase 
behavior of the biaxial particles as a function of the cross-section geometry.
For low values of the aspect ratio of the particle cross section, we
obtain smectic phases with tetratic symmetry, although metastable with 
respect to the crystal, as our MC simulation study indicates. 
For large particle aspect ratios and in analogy with previous work 
[Phys. Chem. Chem. Phys. {\bf 5}, 3700 (2003)], we have 
found a four--phase point where four spinodals, corresponding to  
phase transitions between phases with different symmetries, meet together. 
The location of this point is quite sensitive to particle 
cross section, which suggests that optimizing the particle geometry 
could be a useful criterion in the design of colloidal particles
that can exhibit an increased stability of the biaxial nematic phase with respect 
to other competing phases with spatial order.
\end{abstract}


\maketitle

\section{Introduction}

The stability of thermotropic 
biaxial nematic phases \cite{Freiser,Nature} has been the subject of many
studies in the last three decades, the driving force for this effort being
their potential for use in display devices and other applications \cite{deGennes}.
Theoretical \cite{Freiser,Alben,Straley,biblio}
and simulation \cite{Allen,Romano,Camp,Berardi} studies 
of a number of one-- and two--component model systems have indicated that
biaxiality is indeed possible in bulk nematic phases, but experimental
identification is difficult and, in fact, very few positive reports have appeared
in the literature \cite{Nature}. 

Biaxial phases are a theoretical possibility in 
fluids made of particles that deviate from a cylindrical shape \cite{Freiser}, something
that all molecules do to a larger or lesser degree. However, a biaxial
particle geometry does not necessarily lead to a biaxial nematic phase: 
the degree of biaxiality has to be large enough, and all interactions have to be
tuned optimally. Some early claims for a clear identification of a biaxial 
phase \cite{Malhete,Chandrasekhar,Praefcke}
were not sufficiently substantiated \cite{Luckhurst}, and it has not been until recently
that biaxial nematics in one--component fluids have been identified
unambiguously \cite{Madsen,Acharya} (although with a very low value of the biaxial
order parameter) in fluids with rigid bent--core (V--shaped) molecules. 
Studies on V--shaped molecules, based on simple statistical models
\cite{Ferrarini,Paulo} and computer simulations \cite{Bates}, 
have provided qualitative theoretical support for the experimental observations. However,
it is difficult to isolate molecular shape as the crucial ingredient that causes biaxiality 
in low--mass {\it molecular} fluids, since specific interactions may play a role.

The model systems on which we focus in the present paper are hard particles, 
which can be thought of as idealised (and quite faithful) 
representations of particle interactions in {\it colloidal} (rather than molecular)
fluids consisting of colloidal particles with a shape anisotropy. In fact, novel
methods to synthesise colloidal metallic particles with a variety of shapes, from
ellipsoids \cite{Wiley} and parallelepipeds to rhombohedra and tetrapods \cite{Murphy,Jun,Pu}, 
are now available, which opens up new avenues for theoretical exploration.
Particularly interesting are particles with rectangular shape (parallelepipeds), since
the presence of sharp edges, corners and flat sides may be the source of new
types of ordering. Metallic nanorods with a rectangular cross section 
have been synthesised using different materials \cite{Sau,Xiang,Sun}, and various theoretical
studies using different techniques and approximations indeed point to complex phase behaviour 
\cite{Martinez-Raton1,Escobedo1,Escobedo2,Escobedo3} for particles with square cross
section.

Biaxial phases in mixtures are also a possibility, as some
theoretical \cite{Stroobants,Vanakaras2,Varga,Wensik} and simulation \cite{Camp2} 
studies demonstrate; a positive identification on an experimental rod--plate system also
exists \cite{Kouwer}. In mixtures, difficulties are associated with competition between
biaxial nematic ordering and nematic demixing transitions \cite{Vanakaras3}.
The stabilisation of phases with partial (smectic phase) or complete 
positional order (i.e. freezing) is certainly an effect that competes strongly
with the formation of a stable biaxial nematic phase in one--component fluids, 
so that the window of particle biaxialities where a biaxial
nematic phase can exist is predicted to be very narrow \cite{Ferrarini}. 
Hard nanorod models seem to be ideal systems to study these problems, since one can
focus just on particle geometry and the effect this has on phase behaviour.

In a previous paper, Vanakaras et al. \cite{Vanakaras} have presented 
theoretical and simulation results for a fluid of parallel hard particles with rectangular 
cross section of arbitrary transverse aspect ratio.  
The results of
Vanakaras et al. indicate that indeed a biaxial nematic phase can be stabilised 
at the expense of the smectic phase when the rectangular cross--sectional aspect 
ratio is sufficiently large. Mixtures of these particles have a considerably
broadened region of stability for the biaxial nematic phase with respect
to the pure--fluid case, which of course implies that mixing is a 
general mechanism to stabilise the biaxial nematic phase in these systems.

In the present paper we revisit and extend the type of particles studied by Vanakaras et al. but 
consider only one--component fluids, again using the approximation that particle interactions
are completely hard. 
Since the Onsager--type theory used by Vanakaras et al. should (by construction)
only provide a gross picture of the phase equilibria, we
propose a sophisticated density--functional theory that overcomes some of
the defficiencies of Onsager--type theories. 
Although our theory can be formulated for a general mixture, we 
particularize here to one--component fluids of particles whose main molecular 
axes are assumed to point along a specified direction (nematic director) but
that possess a general cross section, characterised by a second molecular
axis, that can freely rotate in the plane perpendicular to the nematic director. 
The theory is used, subject to some simplifying assumptions,
to study the stabilisation of nematic and smectic mesophases
for a number of particle geometries having different cross-sectional
areas, such as the rectangular and elliptical, among others.
Our theoretical scheme can therefore assess the relative stability 
between uniaxial and biaxial nematic, and uniaxial and biaxial
smectic phases. 

Our proposal, based on the different phase diagrams presented, is that, by optimising the
particle cross section (i.e. considering a wider range of geometries, not
necessarily rectangular), one can also improve the stability range of the biaxial
nematic phase in one--component fluids made of colloidal nanoparticles.
A universal (i.e. {\it independent} of the type of particle) feature of the phase 
diagrams obtained is that the two (uniaxial and biaxial) nematic and the
corresponding smectic phases (four altogether) meet at a `four--phase point',
also observed by Vanakaras et al. for rectangular cross sections,
resulting from the convergence of the corresponding four second--order
transition lines separating pairs of phases. The location of this point,
which is a stability boundary for the biaxial nematic phase, depends on
the particle geometry and this suggests a
mechanism to enhance this stability.
This result may be relevant for the design and synthesis of colloidal particles 
exhibiting biaxial phases. An additional prediction of our theory is that, for
particles with rectangular section and low transverse aspect ratio, a further, 
tetratic smectic phase, possessing four-fold
symmetry in the transverse plane, appears in the phase diagram, albeit in
metastable form.
This phase is reminiscent of the corresponding tetratic nematic phase observed
in two--dimensional fluids of hard rectangles \cite{Schlacken,Martinez-Raton0,Chaikin}.
In fact, the topology of the phase diagram in the case of rectangular
areas seems to have the same limit as the corresponding hard--rectangle
fluid in two dimensions at high packing fractions, a result that reflects
the dimensional crossover property exhibited by the density functional.

The paper is arranged as follows. In the following section we introduce the
density--functional theory, with relevant details on the numerical implementation 
relegated to the appendix. Sec. \ref{results} contains the results for all
the particle geometries considered, separating the rectangular geometry from
the rest, but stressing the differences and similarities in phase behaviour.
We end with some conclusions and perspectives for future work.

\section{Fundamental--measure functional for parallel particles}
\label{Theory}

The system to be studied consists of a fluid of hard biaxial particles 
with characteristic lengths $L$ (parallel to the $z$ axis),   
$\sigma_1$ and $\sigma_2$ (both perpendicular to the $z$ axis). 
The cross-section (transverse) area of the particle does not vary along the 
long axis of the particle, which is taken to lie along the $z$ axis 
(primary nematic director); particles can otherwise 
freely rotate about this axis. Thus the fluid is described in terms of the density 
profile $\rho(z,\phi)$, with $\phi$ the azimuthal angle of the particle
second axis with respect to a fixed direction in the transverse plane (secondary
nematic director). Since the long
axis is fixed, a trivial scaling along this direction follows, and the phase 
behaviour is not going to depend on $L$; however, we expect a strong 
dependence on the particle transverse aspect ratio 
$\kappa=\sigma_1/\sigma_2$ ($\sigma_1$ being the larger size, along the 
particle second axis). 

\begin{figure}
\epsfig{file=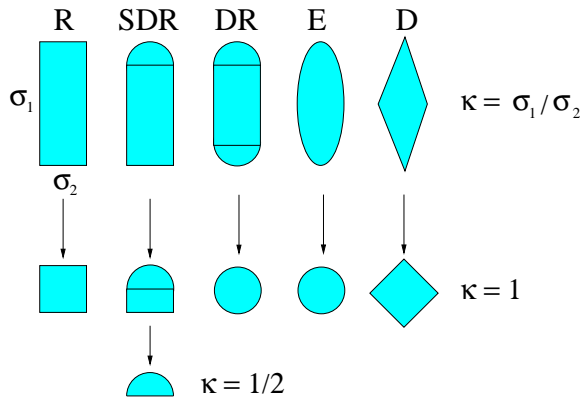,width=3.in}
\caption{(Color online). Sketch of particle transverse sections used in the phase--diagram 
calculations. The limiting case $\kappa=1$ is sketched, and for SDR the 
limiting case $\kappa=1/2$ is also shown.} 
\label{fig1}
\end{figure}

We have chosen five different transverse sections, having
symmetries as shown in Fig. \ref{fig1}: rectangle (R), semi--discorectangle 
(SDR, consisting of a rectangle capped with only one semi--disc in one of their 
ends), discorectangle (DR, obtained from the previous one by adding another 
semi--disc at the other end), ellipse (E), and deltoid (D, composed of
an isosceles triangle and its reflection through its common base). Note that 
both the latter and the rectangular geometries degenerate into a square when 
$\kappa=1$, while ellipses and disco--rectangles both have a disc as a limit.
Finally, the minimum value of aspect ratio for SDR is $\kappa=1/2$, 
corresponding to a semi-circle (see Fig. \ref{fig1}). 

In the following we obtain a fundamental--measure density functional for 
the smectic phase of a fluid composed of biaxial particles with a given 
cross section (in the sense explained above) in the approximation that
the particle long axes are parallel. For this purpose 
we will adopt the \emph{projecting procedure} to construct a functional for
such a three--dimensional fluid starting from the corresponding two--dimensional 
functional for particles with the same section. Details of this formalism are 
given in Refs. \cite{Cuesta} and \cite{Capi}. 

Since we consider the smectic as the less symmetric liquid-crystalline phase
in the present calculations, the corresponding
two-dimensional particles are not assumed to exhibit any spatial ordering in 
the $xy$ (transverse) plane and, therefore, the excess part of the free--energy 
density can be constructed from scaled--particle theory (SPT) [the uniform 
limit of most fundamental--measure functionals] by (i) making the 
density profiles depend only on the $z$--coordinate, and (ii) defining two weighted 
densities, namely the local packing fraction, 
\begin{eqnarray}
\eta(z)=a\int_{z-L/2}^{z+L/2}\rho(z')dz',
\label{packing}
\end{eqnarray}
where $\rho(z)=\int d\phi\rho(z,\phi)$ and $a$ is the cross-section area
of the particle, and the two--particle weighted density
\begin{eqnarray}
N_2(z)&=&\int_{z-L/2}^{z+L/2}dz_1\int_{z-L/2}^{z+L/2}dz_2\int_0^{2\pi} d\phi_1
\int_0^{2\pi}d\phi_2\nonumber\\
&\times&\rho(z_1,\phi_1)\rho(z_2,\phi_2)A(\phi_{12}),
\label{N2}
\end{eqnarray}
where $\phi_{12}=\phi_1-\phi_2$ and 
we defined $A(\phi)\equiv A_{\rm{exc}}(\phi)/2-a$, with 
$A_{\rm{exc}}(\phi)$ the excluded area between two cross sections. 
Thus, we obtain the following expression for the generating function of 
the excess free--energy density \cite{Cuesta}:
\begin{eqnarray}
\tilde{\Phi}(z)=-\frac{\eta(z)}{a}
\ln\left[1-\eta(z)\right]+\frac{N_2(z)}{1-\eta(z)}.
\end{eqnarray}
Using the \emph{projecting procedure} \cite{Cuesta,Capi}, the excess part of the 
three--dimensional free--energy density can be obtained
using the formula
\begin{eqnarray}
\Phi(z)=\frac{\partial}{\partial L}\tilde{\Phi}(z),
\end{eqnarray}
resulting in 
\begin{eqnarray}
\Phi(z)&=&-n(z)\ln\left[1-\eta(z)\right]+\frac{n(z)\eta(z)+N_1(z)}{1-\eta(z)}
\nonumber\\ &+&\frac{ an(z)N_2(z)}{\left[1-\eta(z)\right]^2},
\label{cero}
\end{eqnarray}
where the one--body weighted density $n(z)$ is defined as
\begin{eqnarray}
n(z)=\frac{1}{2}\left[\rho\left(z-\frac{L}{2}\right)+\rho\left(z+\frac{L}{2}\right)\right],
\label{one-body}
\label{dos}
\end{eqnarray}
while a new two--particle weighted density is obtained as 
\begin{eqnarray}
&&N_1(z)=\int_{z-L/2}^{z+L/2}dz_1\int_0^{2\pi} d\phi_1\int_0^{2\pi} d\phi_2
\rho(z_1,\phi_1)\nonumber\\
&&\times\left[\rho(z-L/2,\phi_2)+\rho(z+L/2,\phi_2)\right]
 A(\phi_{12}).
\label{N1}
\end{eqnarray}
The excess part of the free--energy functional for the smectic phase 
is then obtained as 
\begin{eqnarray}
\frac{\beta {\cal F}_{\rm{ex}}}{V}=d^{-1}\int_0^d dz \Phi(z),
\end{eqnarray}
with $\Phi(z)$ 
given by Eqn. (\ref{cero}), $d$ being the smectic period. 
The ideal part is given by 
\begin{eqnarray}
\frac{\beta {\cal F}_{\rm{id}}}{V}=d^{-1}\int_0^d dz\int_0^{2\pi} d\phi 
\rho(z,\phi)\left[\ln \rho(z,\phi){\cal V}-1\right],
\end{eqnarray}
with ${\cal V}$ the thermal volume. This ends the description of the theoretical
tools that will be used to study the phase behavior of biaxial particles with 
different geometries. The equilibrium state of the system follows from 
minimisation of the total free energy density ${\cal F}=\left(
{\cal F}_{\rm{id}}+{\cal F}_{\rm{ex}}\right)/V$. 
The minimisation will be performed numerically,
using a variational procedure, and 
adopting two different approximations for the parameterizations of the 
smectic density distribution (both of which obviously contain the correct
one--particle distribution of the higher--symmetry phases). These parameterizations
are described in detail in Appendix \ref{AI}.

\section{Results}
\label{results}

This section is divided into two parts.  In Sec. 
\ref{rectangles} we describe in detail the results obtained for a system of 
hard biaxial parallelepipeds (rectangular cross section), while in Sec. 
\ref{todo} we present the phase diagrams obtained for the other cross sections,
stressing the most important differences in phase behavior.
The phases found in the phase diagrams and the notations used are: uniaxial 
nematic (N), biaxial nematic (N$_{\rm{B}}$), tetratic nematic (N$_{\rm{T}}$), 
uniaxial smectic (Sm), biaxial smectic (Sm$_{\rm{B}}$), and
tetratic smectic (Sm$_{\rm{T}}$). As shown later, 
the main conclusion that can be drawn 
for this study is that the variation of the cross--sectional geometry has 
a dramatic impact on the relative stability of the N$_{\rm{B}}$ 
phase. This in turn suggests a relatively simple criterion, useful in the
design of colloidal particles, to enhance the stability 
of the biaxial nematic phase with respect to non--uniform phases
(such as the different smectic phases considered here). The underlying mechanism is 
alternative to that observed by Vanakaras et al. \cite{Vanakaras}, where an increase 
of the N$_{\rm{B}}$ stability follows by mixing two species with different sizes.
 
\subsection{Hard biaxial parallelepipeds}
\label{rectangles}
Here the cross section is a rectangle with aspect ratio $\kappa=
\sigma_1/\sigma_2$. We begin with a comparison between our theoretical model and 
standard isobaric Monte Carlo (MC) simulations conducted on systems of $N\simeq 10^3$ 
biaxial parallelepipeds with their long axes parallel to the $z$ axis. These systems
require long times to equilibrate, so simulations in excess of $2\times
10^6$ sweeps
per particle are needed. The main goal here is to make the comparison at the level
of the equation of state (EOS), and for this purpose we chose $\kappa=2$,
i.e. a small value of the aspect ratio (since this is a harder test for the 
theory than a large value). Minimisation of the functional 
was done using the decoupling approximation (see Sec. \ref{decoupling}), which gives exact 
results for the simple Sm symmetry, with no in--plane orientational ordering,
but is only approximate when in--plane order builds up. As we cover the N and Sm phases,
ranging from small to high densities (but always below the Sm-Sm$_{\rm{T,B}}$ transition), 
we have used the Gaussian parameterization proposed in Sec. \ref{Gaussian}. 

\begin{figure}
\epsfig{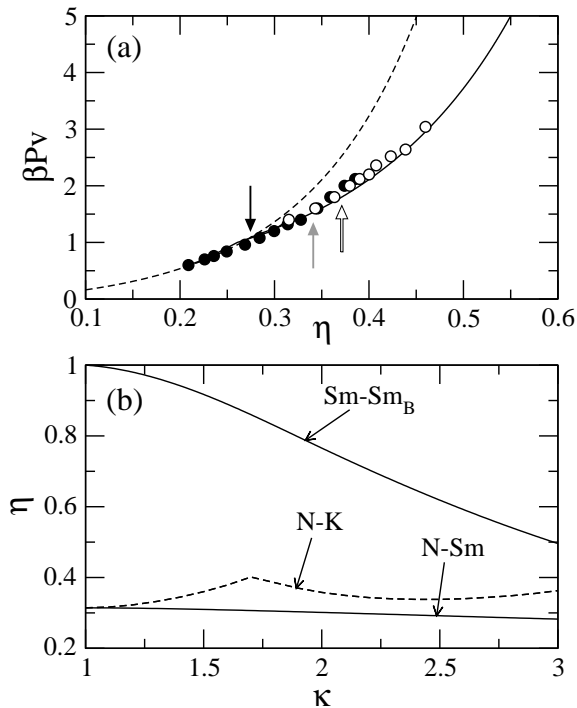}
\caption{(a) Theoretical equations of state of the N (dashed curve) 
and Sm (solid curve) phases of hard biaxial parallelepipeds with $\kappa=2$. 
Results from MC simulations are shown with symbols. Filled circles:
compression run from nematic phase; open circles: expansion run from perfect 
biaxial crystal
at high pressure. Bifurcation points of the N-Sm second--order transition 
obtained by theory and simulation are indicated by filled and shaded arrows, 
respectively. The 
open arrow indicates the approximate Sm-crystal transition 
as obtained by simulation (see text). 
(b) Phase diagram obtained from
the Zwanzig approximation for small values of $\kappa$. Solid curves:
N-Sm and Sm-Sm$_{\rm{B}}$ spinodals; dashed curve: N-K spinodal.}
\label{fig1a}
\end{figure}

The resulting EOS, pertaining to the N 
and Sm branches, is shown in Fig. \ref{fig1a}(a),  
along with the simulation results. First, we comment on the latter.
A compression run (filled circles) was performed from the low-density
nematic phase. At $\eta\simeq 0.31$ small--amplitude density
waves began to develop, which gave rise to a fully developed smectic
density distribution at $\eta\simeq 0.34$ (the `pretransitional'
modulation is probably due to the small system size along the $z$ direction). 
The smectic structure exhibits
no in--plane order of any kind (either translational or orientational), and 
thus can be identified as a standard Sm phase. However, at $\eta\agt 0.37$ 
some kind of translational order sets in, resulting in defected density 
distributions along $z$, still without any orientational order in 
the $xy$ plane. These structures may result from a tendency of the
system to develop crystalline order; the fairly low value of
packing fraction at which this phenomenon occurs suggests that a
plastic solid phase (with particles located at the nodes of 
a 3D lattice, but with their second axes randomly oriented) may be involved.
An expansion run (open circles in Fig. \ref{fig1a}(a) from 
a perfect (biaxial) crystal at high packing fraction also produces such
defected structures and does not help to clarify the situation. However,
the suggestion that a plastic solid might be stabilised is indirectly 
supported by theoretical calculations of the
spinodal to a crystal phase (K), to be presented below.
The main conclusion from the present simulation study is that the 
window of smectic stability, $\Delta\eta\simeq 0.03$, is relatively small 
for moderate aspect ratios.
Further study is required to obtain a more quantitative picture of the
phase behaviour of this system in the crystal region.
In any case, one can see that the
comparison between theory and simulation is fair, as far as the value of the
pressure is concerned. The location of the N--Sm spinodal point as predicted 
by simulations is $\eta\simeq 0.34$ while the theory gives a 
value of $\eta=0.274$; this is reasonably close to the simulation result.  

Our theory does not make any prediction on the transition to a crystalline
phase; in fact, the extension of the present model 
to include columnar or crystalline ordering is not a trivial task. Even if 
an approximate functional could be proposed, its numerical minimisation would
require huge numerical work. Therefore, in an effort to elucidate
the system behaviour observed in the simulations for $\eta\agt 0.37$, 
we have performed a stability analysis in the framework of the
restricted--orientation approximation (Zwanzig approach) to estimate 
the location of the N--K phase transition. The approximation 
involves a constraint on the orientation of the particle second axis to
lie parallel to either the $x$ or $y$ axes. In this context, a 
fundamental--measure
density functional was obtained in Ref. \cite{Cuesta}, which can be
applied to study phases with any spatial symmetry, in particular the
crystal. 

Fig. \ref{fig1a}(b) is the phase diagram for hard parallelepipeds 
with small aspect ratios (between 1 and 3),
as obtained from the Zwanzig approach.
The continuous curves are the N-Sm and Sm-Sm$_{\rm{B}}$
spinodals, while the dashed curve is the spinodal instability in the N phase
with respect to crystal fluctuations; these fluctuations are seen to
correspond to a {\it plastic} solid, as suggested by the simulations. 
As can be seen 
from the figure, for the particular case $\kappa=2$, the packing--fraction 
interval between the N-Sm and N-K spinodals is $\Delta\eta=
\eta_{\rm{K}}-\eta_{\rm{Sm}}= 0.3579-0.3013=0.0566$, a result consistent with 
the simulations.  However, this result is to be taken with care, because the 
Sm-K transition is expected to be of first order (since both phases have 
different symmetries) and, consequently, the bifurcation analysis from the N 
to K phase is but a gross estimate for the location of this transition. 
All we can say for certain is that the K phase bifurcates from the nematic
at a packing fraction above (but close to) the N-Sm transition (although it 
is possibly metastable within some density interval after bifurcation).
These results also show that, for small aspect ratios, the Sm$_{\rm{B}}$ 
phase is unstable with respect to the K phase.

\begin{figure}
\epsfig{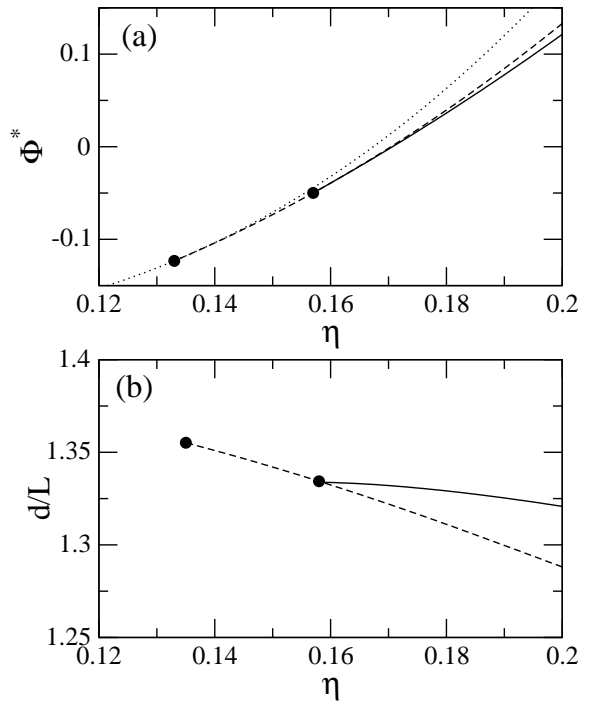}
\caption{(a): Free energy density $\Phi^*=\beta {\cal F} v/V$
vs. packing
fraction $\eta$ of the N (dotted curve), Sm (dashed curve) and Sm$_{\rm{B}}$
(solid curve) phases of particles with R cross section and with aspect
ratio $\kappa=15$. (b): smectic period of Sm (dashed curve)
and Sm$_{\rm{B}}$ (solid curve) phases.
The symbols represent the N-Sm and Sm-Sm$_{\rm{B}}$ bifurcation
points, respectively.}
\label{fig2}
\end{figure}

For higher aspect ratios (for which we have not performed simulation studies), 
in particular for $\kappa=15$, we have found that the Sm$_{\rm B}$ phase
is stabilized at low values of the packing fraction. To obtain the phase 
behavior for this particular case, we have taken advantage of the fact that
the mean density is relatively small and, therefore, we have used the 
Fourier transform parametrization of Sec. \ref{Fourier}, 
which gives a quasi--exact representation of the true density profile
(numerical convergence is guaranteed in this regime of $\eta$). The results are shown 
in Fig.  \ref{fig2} (a), where the free--energy density of the N, Sm and 
Sm$_{\rm{B}}$ phases is plotted as a function of the packing fraction. The system 
exhibits a second--order N--Sm transition, a relative small window of Sm stability, 
and then a continuous Sm--Sm$_{\rm{B}}$ transition, the latter being the 
stable phase for higher densities (up to the freezing transition). Thus, the 
two--dimensional orientational ordering of the particle second axis appears in a 
continuous fashion with increasing density. In Fig. \ref{fig2} (b) 
the smectic period is plotted as a function of packing fraction for both types
of smectic phases. It is interesting to note that the period of the 
Sm$_{\rm{B}}$ 
phase decreases very slowly with density, compared with the 
corresponding behaviour in the Sm phase. 

\begin{figure}
\epsfig{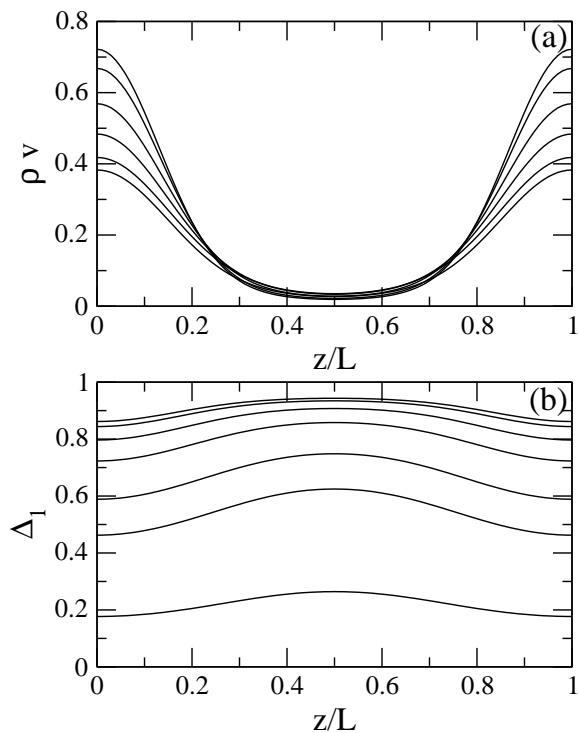}
\caption{Evolution of the density profile (a) and biaxial order parameter (b)
of the smectic phase of particles with R cross section and aspect ratio
$\kappa=15$ for $\eta=0.16 +0.02i$ ($i=0,\cdots,4$) and
$\eta=0.25$ (from bottom to top). In (b) the case $\eta=0.17$ is also included.}
\label{fig3}
\end{figure}

With the aim to understand the non--uniform spatial and orientational 
correlations in the Sm$_{\rm{B}}$ phase, we have plotted in Fig. \ref{fig3}   
the evolution of the density $\rho(z)$ 
and the biaxial order parameter (see \ref{Fourier}) $\Delta_1(z)$ profiles with 
the mean packing fraction $\eta$. While the density inhomogeneities 
build up with packing fraction,  
the order parameter profile, although globally increasing, becomes flat as
a function of  
$z$ with increasing $\eta$. Also, the profiles are out of phase, 
i.e. particles at smectic layers have a slightly
lower orientational order than those situated 
at the interstitials. This is an interesting structural feature that points to
a non--trivial coupling between layers via interstitial particles.
However, this weak effect becomes less and less relevant as $\eta$ 
increases, as revealed by the function $\Delta_1(z)$ becoming practically constant as a 
function of $z$. The latter fact justifies a posteriori the use of the 
decoupling approximation to study the Sm$_{\rm{B}}$ phase and, probably, also other  
phases, such as the Sm$_{\rm{T}}$ phase that appears at higher densities and small 
values of $\kappa$.

\begin{figure}
\epsfig{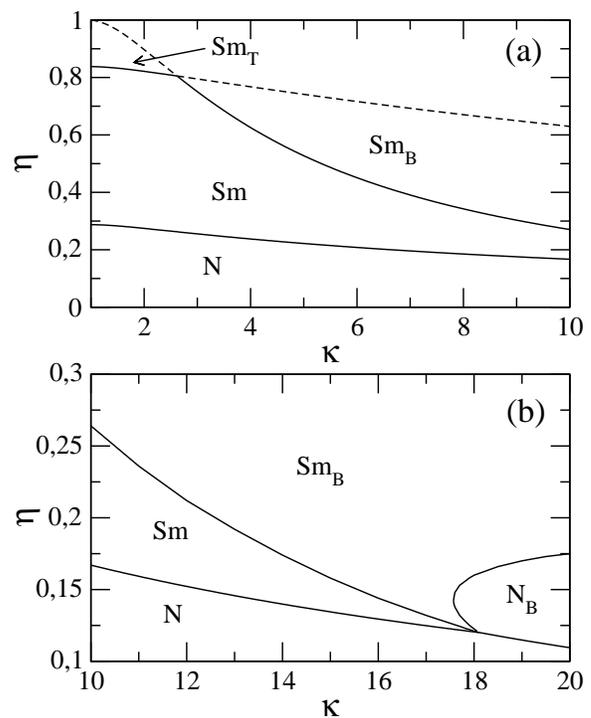}
\caption{Phase diagram for biaxial parallelepipeds 
(R cross section) calculated using the decoupling approximation
with the Gaussian parametrization (a) and the
Fourier expansion (b) for the density profile.
The solid curves represent the continuous
phase transitions between phases with different symmetries, as labelled in
the figure. In (a) the dashed
curves show the continuation of the Sm-Sm$_{\rm{B}}$ 
and Sm-Sm$_{\rm{T}}$ spinodals.}
\label{fig4}
\end{figure}

To calculate the global phase diagram we have used two approximations:
the decoupling approximation $\rho(z,\phi)=\rho(z)h(\phi)$ 
with a Gaussian parameterization of $\rho(z)$ for $\kappa\leq 10$, and 
the Fourier--transform parametrization of the complete function $\rho(z,\phi)$ 
for $\kappa>10$. This choice is motivated by the 
dependence of the numerical criterion for convergence, in the minimization routines,
on the value of $\eta$. The phase diagram is plotted in Figs. \ref{fig4} (a) and (b). For small 
aspect ratios, $1\leq \kappa\leq 2.618$, we find a N--Sm transition at low
densities, the Sm phase being stable up to $\eta \simeq 0.8$, beyond which the fluid
exhibits a continuous transition to a Sm$_{\rm{T}}$ phase. This transition was 
calculated using Eq. (\ref{spin0}) with $i=2$. As described previously, 
the Sm$_{\rm{T}}$ phase consists of smectic layers in which the second particle 
axes (parallel to the smectic planes) point, with equal probability, along
two mutually perpendicular directions (secondary nematic directors); within  
our approximation, the orientational distribution function fulfills the 
symmetry $h(\phi)=h(\phi+\pi/2)$. This phase is sandwiched between the Sm and 
the Sm$_{\rm{B}}$ phases. Although we have not calculated numerically the location
of the Sm$_{\rm{T}}$-Sm$_{\rm{B}}$ transition, it can be approximated, as
noted in a previous work \cite{Martinez-Raton0}, by the 
N--Sm$_{\rm{B}}$ spinodal extended to small values of $\kappa$ 
[the dashed curve of Fig. \ref{fig4} (a)]. 

The Sm-Sm$_{\rm{T}}$ phase transition 
can be preempted by a transition to a crystalline phase
with tetratic symmetry. Nevertheless, we have shown 
the high density part of the phase diagram with only one-dimensional periodic phases
included. As we will see later, these results are very useful for 
testing the performance of the present functional in the description of 
highly inhomogeneous fluids. Further, for aspect ratios in the range 
$2.618\leq \kappa\leq 18.101$, the window of Sm stability (between the N 
and the Sm$_{\rm{B}}$ phases) decreases with $\kappa$, disappearing altogether at
$\kappa=18.101$, the point where the N--Sm and Sm--Sm$_{\rm{B}}$ spinodals meet. 
This point was calculated using Eqns. (\ref{spin})--(\ref{sistema}), with 
the inverse of the structure factor given by Eqn. (\ref{structure}). For higher 
values of $\kappa$ the N phase exhibits a second--order transition to 
the N$_{\rm{B}}$ phase at a packing fraction calculated through Eqn. 
(\ref{spin}). On further increasing the density, there is a continuous 
transition between the N$_{\rm{B}}$ and Sm$_{\rm{B}}$ phases. The 
N$_{\rm{B}}$--Sm$_{\rm{B}}$ and the N--N$_{\rm{B}}$ spinodals meet 
at the point mentioned in the introduction, 
which will be called {\it four--phase} point
\cite{Vanakaras} since four different spinodals meet at the same point in
the phase diagram. It is interesting to note that the transition between the
N$_{\rm{B}}$ and Sm$_{\rm{B}}$ phases is reentrant in an interval of
aspect ratios just below the four--phase point. 
This is a genuine prediction of our theory, since
Onsager theory predicts a monotonic phase boundary between
these two phases \cite{Vanakaras}.
 
It should be noted that our prediction for the location of the four--phase point at 
$\kappa=18.101$, using our density--functional approximation, is to be contrasted
with the value $\kappa\sim 15$ reported in Ref. \cite{Vanakaras}, where a second--virial 
Onsager theory was used instead. The MC simulations carried out in Ref. \cite{Vanakaras}
seem to confirm this latter value. However, this apparent agreement should be taken
with some caution. First, Onsager theory is known to give a poor picture of the 
N--Sm transition due to the misrepresentation of density correlations. Also, excluded--volume
effects underlying in--plane orientational ordering are probably not enough to give
a quantitative description, since higher--than--two--body correlations are known to 
be very important in two dimensions, and the problem at hand, once smectic layers
have been formed, is quasi two--dimensional in nature. On the other hand, simulations
of these systems are difficult, and large system--size effects are expected.
The agreement found in Ref. \cite{Vanakaras} could be just fortuitous. Our approach,
which includes higher--order correlations, should in principle give a more representative 
picture, but the situation is difficult to assess for lack of more extensive computer
simulations and theoretical studies. For example, our theory gives only an approximate value for the 
third virial coefficient and, as shown in Ref. \cite{three-body}, this 
coefficient is of the same order of magnitude 
as the second one for two-dimensional 
particles (the particle cross sections) in the the limit $\kappa\to\infty$. 
Thus, for high aspect ratios 
our theory can deviate from the Onsager theory. A third--virial theory, 
including the exact virial coefficients up to the third order, is required to 
improve understanding of this issue.

\subsection{Other cross sections}
\label{todo}

In this section we present the phase diagrams corresponding to the other
particle cross sections described at the beginning of Sec. \ref{Theory}. 
As the symmetries of the different phases and the nature of their phase transitions 
were discussed in the preceding section, here we will concentrate only on describing
the differences between the phase diagrams of the different particle
geometries. 

\begin{figure}
\epsfig{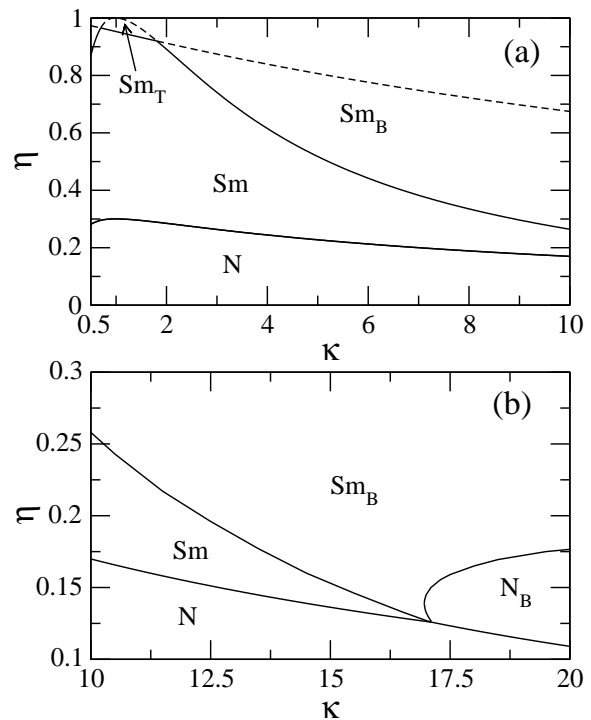}
\caption{Phase diagram for particles with SDR cross section. 
Labels and lines are as in Fig. \ref{fig4}.}
\label{fig5}
\end{figure}

\begin{figure}
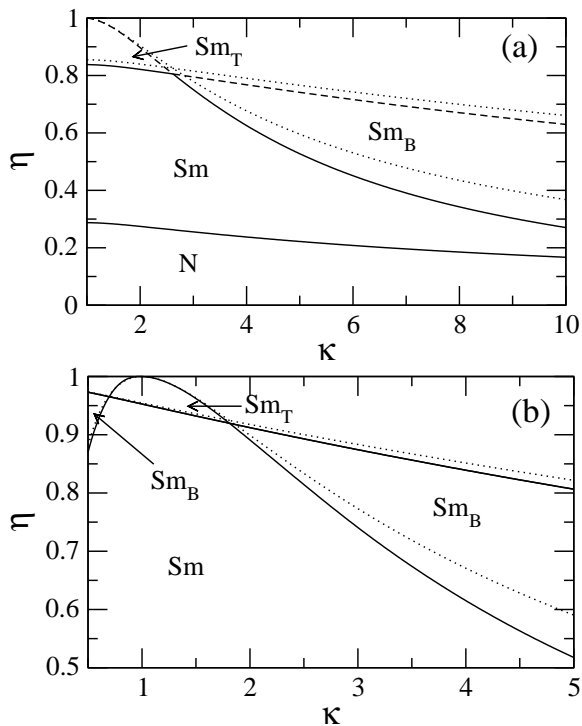

\epsfig{file=Fig7a.eps,width=3.in}
\epsfig{file=Fig7b.eps,width=3.in}
\caption{Phase diagrams for R (a) and SDR (b) particles 
in the neighborhood of the
Sm$_{\rm{T}}$ phase. All transition lines for
three-dimensional fluids are shown with solid curves, while those 
corresponding to two-dimensional \cite{Martinez-Raton0}
with the same particle transverse sections are shown with dotted curves. 
For 2D systems one needs to make the substitution $\text{Sm}\to\text{I}$ 
(isotropic), 
$\text{Sm}_{\rm{B}}\to\text{N}$ (uniaxial nematic), 
and $\text{Sm}_{\rm{T}}\to\text{N}_{\rm{T}}$ 
(tetratic nematic).}
\label{fig6}
\end{figure}

We begin by presenting the results corresponding to particles 
with SDR cross section (this is the only particle that does not possess 
head--tail symmetry). In Figs. \ref{fig5} (a) and (b) we show the phase diagram. 
The location of the four--phase point, $\kappa=17.121$, is shifted to lower 
values of the aspect ratio, compared with the preceding case (rectangular cross section). Also, an important 
difference lies in the phase behaviour for small $\kappa$ and high packing fraction. To
better visualise the difference, a zoom is shown in Figs. \ref{fig6} (a) and (b) in 
the region of Sm$_{\rm{T}}$ stability corresponding to R and 
SDR cross-sections, respectively. 

It is apparent from the figures that the stability of the Sm$_{\rm{T}}$ phase 
($0.696\leq \kappa\leq 1.810$) decreases by adding a semi--disc at one end of the 
rectangular section. This can be explained by the increasing excluded volume involved
in the T--configuration (two particles in a perpendicular configuration) when this 
semidisc is added. It is also interesting to note that,
for $0.5\leq\kappa\leq 0.696$, 
a transition between the Sm and Sm$_{\rm B}$ phases again occurs, due to the fact that the 
rectangular part is so small compared with the total area of cross section that
the T--configuration is not entropically favoured. Finally, in Figs. \ref{fig6} (a) and (b), 
the spinodals of the transitions between the isotropic phase (I) and the nematic
and tetratic nematic phases of a strictly two--dimensional fluid
of particles, with the same cross sections as analyzed here, are plotted.
A relevant conclusion that can be drawn from the figure is the coalescence of 
these spinodals and those of the three--dimensional particles
as packing fraction increases. This result confirms the dimensional 
cross--over property of the present density functional. For high $\eta$
the smectic phase can be considered as a collection of smectic layers where 
particles are perfectly located, and the effective interaction between 
particles located at different planes play a secondary role.

\begin{figure}
\epsfig{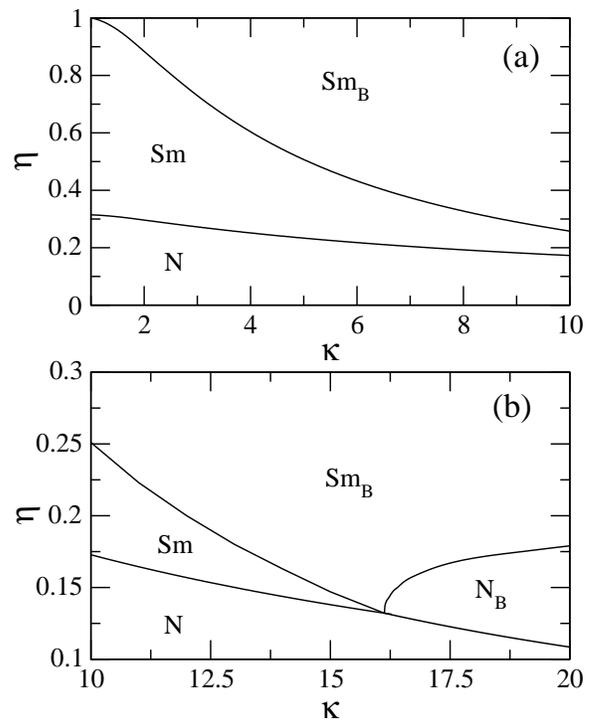}
\caption{Phase diagram of particles with DR cross section. 
Labels and lines are as in Fig. \ref{fig2}.}
\label{fig7}
\end{figure}

\begin{figure}
\epsfig{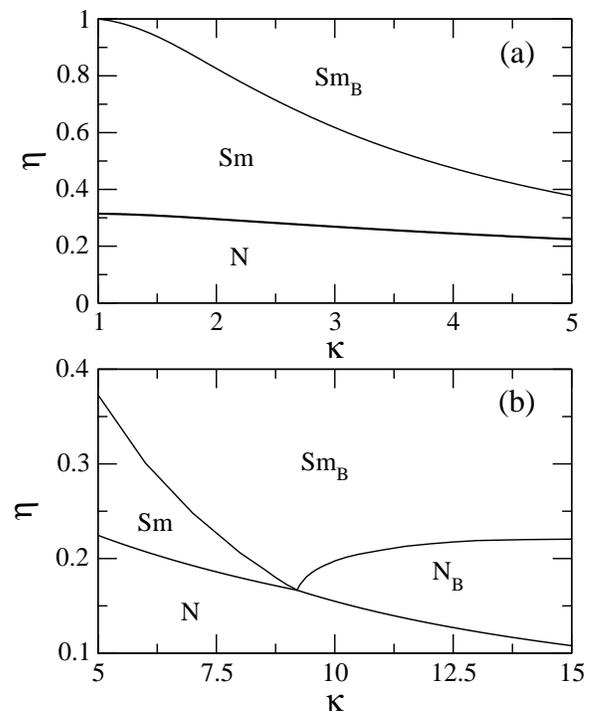}
\caption{Phase diagram of particles with E cross section. 
Labels and lines are as in Fig. \ref{fig2}.}
\label{fig8}
\end{figure}

\begin{figure}
\epsfig{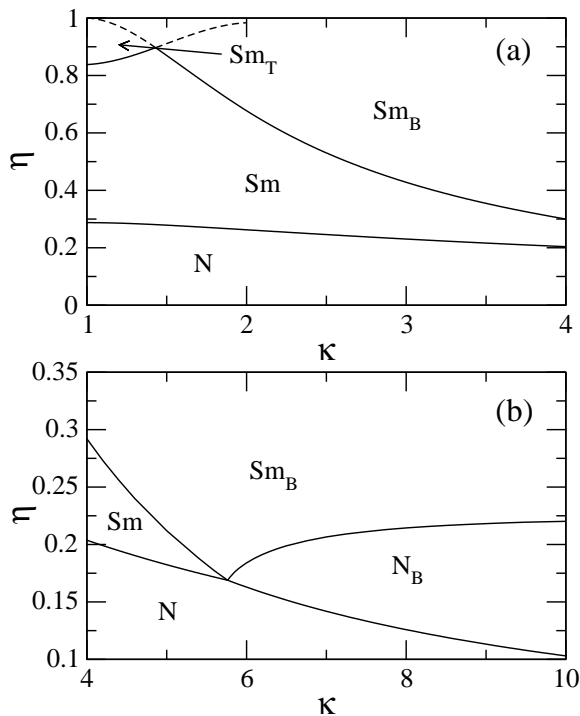}
\caption{Phase diagram of particles with D cross section. 
Labels and lines are as in Fig. \ref{fig2}.}
\label{fig9}
\end{figure}

The phase diagrams for particles with cross sections corresponding to 
DR and E geometries are shown in Figs. \ref{fig7} (a)-(b) and \ref{fig8} (a)-(b), 
respectively. These two phase diagrams have the common feature that 
the Sm$_{\rm{T}}$ phase is absent. The four--phase points are now located at 
$\kappa=16.131$ and $\kappa=9.205$, respectively. Finally, Figs. \ref{fig9}
(a)--(b) are the phase diagrams of particles with D section. Now at high packing fractions, 
and for aspect ratios in the range $1\leq \kappa \leq 1.430$, the Sm phase 
exhibits a second--order transition to a Sm$_{\rm{T}}$ phase. However, the spinodal 
for this transition [$\eta(\kappa)$] is an increasing function of $\kappa$, which 
can be explained by the change of particle geometry with $\kappa$. 
The deltoid in the limit $\kappa=1$ coincides with a square;
departure from this limit by increasing $\kappa$ involves a change in
the angle between the adjacent sides of the deltoid from its initial value of 
90$^{\circ}$ and, as a consequence, the T--configuration of a pair of particles 
(which is characteristic of the tetratic symmetry) is less favoured.    

An additional feature that depends on the particle geometry is the
occurrence of reentrant behaviour in the transition between the
N$_{\rm B}$ and Sm$_{\rm B}$ phases. This behaviour is clearly 
associated with the rectangular nature of the particle shape,
as it only appears in the phase diagrams of R and SDR particles,
and much more pronounced in the former. Since both phases
exhibit biaxial order, and the reentrant transition
involves nematic to smectic ordering (or vice versa), the effect
is the result of a nontrivial coupling between spatial ordering
along the $z$ direction and angular ordering in the transverse plane.

\begin{figure}
\epsfig{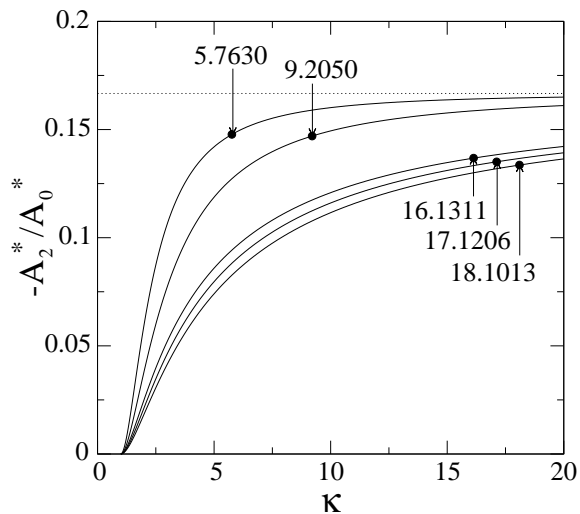}
\caption{Ratio $-A_2^*/A_0^*$ as a function of $\kappa$ for different 
particle geometries corresponding to R, SDR, DR, E and D (from bottom 
to top). The location of the four--phase points are also shown (filled circles). The dotted 
curve represents the asymptote $\kappa\to\infty$ for all the curves shown.}  
\label{fig10}
\end{figure}

To explain the evolution of the four--phase points with 
the change in particle geometry we resort to Fig. \ref{fig10},
where the ratio between the coefficients $-A_2^*$ and 
$A_0^*$ is plotted for different cross sections. This ratio 
is a measure of the relative reduction in excluded volume, or relative gain
in free volume, when the particles are orientationally ordered along the 
nematic director. As can be seen from the figure, this gain increases by modifying the 
particle sections in the sequence R, SDR, DR, E and D. This in turn
explains the sequence found in the location of the four--phase point, namely
the N$_{\rm{B}}$ phase appears, for the first time, for the particle geometry that 
maximizes the gain in free volume associated to the orientational ordering of the second 
particle axis. 

\section{Conclusions}
In the present article we have analysed the phase behaviour of models of
particles that exhibit biaxial liquid--crystalline order, with an emphasis
on biaxial nematic phases. The models consist of hard particles with their
principal axes parallel to each other, and such that the cross section along
this common axis is constant while the secondary axis associated with this
cross section can otherwise rotate freely in the plane. The positive
identification of a biaxial nematic phase requires that its stability be 
compared with that of competing phases with spatial order, such as the smectic 
phase. In order to incorporate these phases into the theoretical scheme, a 
proper treatment of correlations has to be done. Since Onsager--type theories
present severe defficiencies in this respect, we have developed 
a density functional, based on fundamental--measure theory, which makes
a more appropriate treatment of correlations at high densities. The theory
has been applied to study nematic and smectic phases with different 
orientational symmetries, such as the biaxial and tetratic symmetries,
and the global phase diagrams, for particles with 
five different cross sections. 

As our first and most important result, we have obtained the evolution 
in phase behavior with particle geometry. For small aspect ratios and for R, 
SDR, DR and D--type sections, we have found a smectic phase with tetratic 
symmetry. 
The spinodals of the phase transitions between Sm and Sm$_{\rm{B,T}}$ phases, 
at high packing fractions, are similar to those corresponding 
to phase transitions between isotropic and uniaxial or tetratic nematic 
phases of a strictly 2D fluid 
composed of particles with the same cross section. 
This result confirms the 
dimensional cross--over property of the functional. However, the 
Sm$_{\rm{T}}$ phase is preempted by the crystalline phase, as the MC simulations 
seem to show. All the phase diagrams for large values of $\kappa$ have a
common feature, consisting in the presence of a four--phase point at which
the four spinodals corresponding to the second--order N--Sm, N--N$_{\rm{B}}$, 
Sm--Sm$_{\rm{B}}$ and N$_{\rm{B}}$--Sm$_{\rm{B}}$ transitions meet. 
By studying the location of this point as a function of particle geometry,
we have obtained a procedure to increase the N$_{\rm{B}}$--phase
stability, which might be useful in the design and synthesis 
of colloidal particles exhibiting a transition to this phase. 
In particular, the deltoid seems to be the cross section that favours the 
N$_{\rm{B}}$ stability most. Note that this idea is alternative to that 
proposed in Ref. \cite{Vanakaras}, which consists in mixing two species with 
the same cross section but different particle lengths.

The density functional proposed in this work can be used in a variety of
situations, in particular, to study interfacial problems and the
effect of confinement (for example 
in slit geometry) on the stability of different smectic phases of a fluid 
composed by biaxial particles. This study we leave for future work.


\begin{acknowledgments}
Y.M.-R. gratefully acknowledges financial support from Ministerio de 
Educaci\'on y Ciencia (Spain) under a Ram\'on y Cajal research contract and 
the MOSAICO grant. This work is part of the research projects 
Nos. FIS2005-05243-C02-01 and FIS2007-65869-C03-01, also from Ministerio 
de Educaci\'on y Ciencia, and grant No. S-0505/ESP-0299 from Comunidad 
Aut\'onoma de Madrid (Spain). Support from the Spanish--Hungarian
`Integrated Actions' programme under grant Nos. HH-2006-0005 
is also acknowledged.
\end{acknowledgments}

\appendix
\section{Density profile parameterizations}
\label{AI}
In \ref{decoupling} we describe the decoupling approximation and the corresponding
expressions for free energy and structure factor. In \ref{Gaussian} a variational 
density profile, based on Gaussian trial functions, 
is proposed which, together with the decoupling approximation,
allows to calculate the high density region of the phase diagrams. 
Finally, in \ref{Fourier} a truncated Fourier expansion of the density profile 
is introduced; this parameterization is nearly exact for the 
description of smectic phases, but has the pitfall that it can only be used 
for low values of the mean densities due to the poor numerical convergence of the 
minimization procedure.

\subsection{Decoupling approximation}
\label{decoupling}

We adopt the usual decoupling approximation for the density profile:
\begin{equation}
\rho(z,\phi)=\rho(z) 
h(\phi),
\label{dec}
\end{equation}
with $h(\phi)$ the orientational distribution function. 
As its name indicates, this approach decouples spatial and 
angular variables, which implies $z$--independent orientational order in the
smectic phase. In turn this means that the biaxial order parameters, defined by
\begin{eqnarray}
\Delta_i=\langle \cos (2i\phi)\rangle\equiv 
\int_0^{2\pi} d\phi \cos(2i\phi)h(\phi), 
\end{eqnarray}
with $i=1$ for uniaxial and $i=2$ for two--dimensional tetratic symmetries, 
respectively, are constant within a smectic period.

Inserting (\ref{dec}) into (\ref{cero}), we obtain
\begin{eqnarray}
\Phi(z)&=&-n(z)\ln\left[1-\eta(z)\right]+
\frac{n(z)\eta(z)\left(1+2\langle \langle A^* \rangle\rangle\right)}
{1-\eta(z)}\nonumber\\
&+&\frac{n(z)\eta(z)^2\langle \langle A^* \rangle\rangle}
{\left[1-\eta(z)\right]^2},
\end{eqnarray}
where the double angular average  
\begin{eqnarray}
\langle\langle A^*\rangle\rangle=\int_0^{2\pi} d\phi_1\int_0^{2\pi} d\phi_2 h(\phi_1) 
h(\phi_2) A^*(\phi_1-\phi_2), \nonumber\\
\end{eqnarray}
has been defined. Also, the dimensionless quantity $A^*(\phi)=A(\phi)/a$
was introduced. Using the Fourier expansion of the orientational distribution 
function,
\begin{eqnarray}
h(\phi)=\frac{1}{2\pi}\left[1+\sum_{k\geq 1}h_{k}\cos(k\phi)\right],
\end{eqnarray}
we obtain, for the double angular average, 
\begin{eqnarray}
\langle\langle A^*\rangle\rangle=\sum_{k\geq 0}A_{k}^*h_k^2,
\end{eqnarray}
with $h_0=1$. The Fourier coefficients $A_k^*$ are given in Appendix \ref{AA}
for all the particle sections studied (note that $A_{2n+1}^*=0$ 
for all the geometries, except for SDR, the only one that breaks 
the head--tail symmetry; see discussion on the consequence of
this in Sec. \ref{Fourier}). 

The continuous N--Sm transition 
can be calculated from the divergence of the inverse 
structure factor $S^{-1}(q,\eta)=1-\rho\hat{c}(q,\eta)$, calculated 
from the Fourier transform of the direct correlation 
function $\hat{c}(q,\eta)$; the latter is 
obtained from the second functional derivative of $\beta {\cal F}$ 
with respect to the density profile. Within the decoupling approximation, this 
results in  
\begin{eqnarray}
&&S^{-1}(q,\eta)=1+2yj_1(q^*)\left[2+y+2(1+y)^2\langle\langle A^*
\rangle\rangle\right]\nonumber\\
&&+j_1(q^*/2)^2
y^2\left[3+2y+6(1+y)^2\langle\langle A^*
\rangle\rangle\right],
\label{structure}
\end{eqnarray} 
with $q^*=qL$, $y=\eta/(1-\eta)$, $j_1(x)=\sin(x)/x$, and 
where the double angular average $\langle\langle A^*\rangle\rangle$ is to be 
evaluated with $h(\phi)=1/(2\pi)$, which gives the coefficient $A_0^*$. 
The equation $S^{-1}(q,\eta)=0$, together with 
$\partial S^{-1}(q,\eta)/\partial q=0$, must be solved for $\eta$ and 
$q$ at the absolute minimum of $S^{-1}(q,\eta)$. 

\subsection{Gaussian parametrization}
\label{Gaussian}
We adopt the following parameterized density profile: 
\begin{eqnarray}
\rho(z)=\rho d \left(\frac{\alpha}{\pi}\right)^{1/2}\sum_{k}
\exp{\left[-\alpha (z-k d)^2\right]},
\label{Gauss}
\end{eqnarray}
i.e. a sum of normalized Gaussian peaks. This normalized form guarantees 
that $d^{-1}\int_0^d dz \rho(z)=\rho$, with $\rho$ the mean smectic density.
Insertion of (\ref{Gauss}) into (\ref{packing}) and (\ref{one-body}) gives
\begin{eqnarray}
n(z)&=&\frac{\rho d}{2}\sqrt{\frac{\alpha}{\pi}}\sum_k
\left\{\exp\left[-\alpha(z+L/2-kd)^2\right]\right.\nonumber\\ &+&\left.
\exp\left[-\alpha(z-L/2-kd)^2\right]\right\},\\
\eta(z)&=&\frac{\eta d}{2}\sum_k
\left\{\text{erf}\left[\sqrt{\alpha}(z+L/2-kd)\right]\right. \nonumber\\
&-&\left.\text{erf}\left[\sqrt{\alpha}(z-L/2-kd)\right]\right\},
\end{eqnarray}
with $\text{erf}(x)$ the standard error function. For smectic
symmetry, and without orientational ordering parallel to the second nematic director,
the expressions for $\langle\langle A^*\rangle\rangle$ 
with $h(\phi)=1/2\pi$ (isotropic distribution function) are analytic functions of the particle 
characteristic lengths and are provided in Appendix \ref{AA} for all the
geometries considered.
We  minimize the resulting free--energy density $\beta {\cal F}(\alpha, d)/V$ with respect 
to the Gaussian parameter $\alpha$ and the smectic period $d$ for a 
fixed mean packing fraction $\eta=\rho aL$. Varying $\eta$ and 
repeating the above procedure, we obtain the free--energy branch for the
Sm phase. The (continuous) nematic--smectic transition is located at
the mean packing fraction value for which $\alpha\sim 0$. Alternatively, 
this transition can be calculated from the divergence of the inverse 
structure factor, defined by (\ref{structure}).   

To calculate the second--order transitions between the Sm and 
the biaxial smectic (Sm$_{\rm{B}}$) or tetratic smectic (Sm$_{\rm{T}}$) 
phases, we have used a bifurcation analysis in which the
orientational distribution function near the bifurcation point is 
approximated as $h(\phi)\approx[1+h_i\cos(2i\phi)]/(2\pi)$ ($i=1$ and $2$ 
for the uniaxial and tetratic symmetries, respectively). After insertion of this 
expression into the free--energy functional $\beta{\cal F}$ obtained 
from the decoupling approximation (see Sec. \ref{decoupling}), we obtain 
the following expression for the free--energy difference per particle
($\varphi=\beta{\cal F}/N$) between the Sm$_B$ (or Sm$_T$) and Sm phases:
\begin{eqnarray}
\Delta\varphi&=&\frac{h_i^2}{4}T(\eta,\alpha^*,d^*;\kappa)=
\frac{h_i^2}{4}\left\{1+\frac{4A_{2i}^*}{\rho d^*}\right. \nonumber\\
&\times&\left.\int_0^{d^*}dz
\frac{n(z)\eta(z)\left[2-\eta(z)\right]}
{\left[1-\eta(z)\right]^2}\right\}.
\label{spin0}
\end{eqnarray}
The spinodal curves ($\eta$ as a function of $\kappa$) are then 
calculated as the solution of the equation $T(\eta,\alpha^*,d^*;\kappa)=0$ 
for $\eta$, where $\alpha^*$ and $d^*$ are those values obtained from 
the minimization of the free--energy density of 
the Sm phase $\beta{\cal F}(\alpha,d)/V$ with respect to the Gaussian 
parameter $\alpha$ and the smectic period $d$.

\subsection{Fourier parameterization and calculation of spinodals}
\label{Fourier}
The density profile is now parameterized by a truncated 
Fourier expansion:
\begin{eqnarray}
\rho(z,\phi)=\frac{\rho}{2\pi}\sum_{k,m\geq 0}^{K,M}
s_{km}\cos(qkz)\cos(2m\phi),
\label{cinco}
\end{eqnarray}
where $q=2\pi/d$ is the wave number. 
The latter, together with the Fourier amplitudes $s_{km}$, span the space of minimization 
variables. The zeroth--component Fourier amplitude 
is set equal to unity, i.e. $s_{00}=1$. Inserting the expression (\ref{cinco}) into the 
definitions of all the one--particle 
weighted densities, Eqns. (\ref{packing}) and (\ref{one-body}), we obtain 
\begin{eqnarray}
n(z)&=&\rho\sum_{k\geq 0}s_{k0}
j_0(qkL/2)\cos(qkz),\\
\eta(z)&=&\eta\sum_{k\geq 0}s_{k0}j_1(qkL/2)\cos(qkz),
\end{eqnarray}
with 
$j_0(x)=\cos x$. Also, we obtain the following 
expressions for the two-particle weighted densities (\ref{N2}) and (\ref{N1}):
\begin{eqnarray}
N_1(z)&=&2\rho\eta\sum_{k_1,k_2,n}s_{k_1n}s_{k_2n}
A_{2n}^*j_0(qk_1L/2)j_1(qk_2L/2)\nonumber\\
&\times&\cos(qk_1z)\cos(qk_2z),\label{seis}\\
N_2(z)&=&\frac{\eta^2}{a}\sum_{k_1,k_2,n}s_{k_1n}s_{k_2n}
A_{2n}^*j_1(qk_1L/2)j_1(qk_2L/2)\nonumber\\
&\times&\cos(qk_1z)\cos(qk_2z),
\label{siete}
\end{eqnarray}
where the coefficients $A_{2n}^*$ for the different geometries are provided in 
Appendix \ref{AA}. The free energy is then minimised with respect to the 
Fourier amplitudes $s_{km}$ and to the wave vector $q$. In practice 
we needed about 50 Fourier components with $K=10$ and 
$M=5$. 

In connection with the Fourier expansion for the density (\ref{cinco}) and the
corresponding expansions for the weighted densities (\ref{seis}) and 
(\ref{siete}), we must note that only coefficients of the excluded area
with {\it even} index, $A^*_{2n}$, have been taken into account. This is 
justified for particles with head--tail symmetry, as $A_{2n+1}^*=0$ for
these particles. However, for SDR particles, which do not
exhibit this symmetry, one has $A_{2n+1}^*\geq 0$. Since, for the nematic 
phase, {\it or} for the smectic phase in the framework of the decoupling 
approximation, we have $\left<\left<A\right>\right>=\sum_{k\geq 
0}A_k^* h_k^2$, the free--energy minimization with respect to the Fourier 
amplitudes always gives $h_{2n+1}=0$. However, if the coupling between spatial 
and orientational degrees of freedom is properly taken into account, 
products of the form $s_{k_1m}s_{k2m}A^*_{m}$, with {\it odd} $m$, do appear 
in the expansions (\ref{seis})--(\ref{siete}) for SDR particles. These
coefficients could be negative, and in principle this could result in 
equilibrium values $h_{2n+1}\neq 0$, i.e. in smectic phases with in--plane 
polar structure in their density profiles $\rho(z,\phi)$. We expect
terms of this type not to be dominant for large $\kappa$, and therefore 
we have neglected these terms in the calculations for SDR particles, 
in the hope that the topology of the phase diagram near the four-phase 
point is not greatly affected by this approximation.

A measure of the local orientational order is given by the order parameters 
\begin{eqnarray}
\Delta_i(z)&=&\frac{1}{\rho(z)}\int_0^{2\pi} d\phi\rho(z,\phi)\cos (2i\phi)\nonumber\\
&=&\frac{\rho}{2\rho(z)}\sum_{k}s_{ki}\cos(qkz),
\end{eqnarray}
with $i=1$ for the uniaxial order and $i=2$ for the tetratic order; here
the relation $\displaystyle{\rho(z)=\rho\sum_ks_{k0}\cos(qkz)}$ should be used.

The second order N--N$_{\rm{B}}$ transition can be calculated using a 
simple bifurcation analysis of the free--energy difference per particle between
the phases, expressed as a truncated power series in
the Fourier amplitudes $h_i$ (retaining only the first term):
\begin{eqnarray}
\Delta \varphi=\frac{h_1^2}{4}\left[1+4A_2^*(2y+y^2)\right].
\label{bifurca}
\end{eqnarray}  
The non-trivial solution to the equation $\partial \Delta \varphi/\partial h_1=0$ gives
\begin{eqnarray}
&&y=\sqrt{1-\left(4A_2^*\right)
^{-1}}-1, \nonumber\\
&& \eta=1-\frac{1}{\sqrt{1-\left(4A_2^*\right)^{-1}}}.
\label{spin}
\end{eqnarray} 
The intersection between the spinodal of the N--Sm transition, calculated using
\begin{eqnarray}
S^{-1}(q,\eta)=
\frac{\partial S^{-1}(q,\eta)}{\partial q}=0, \quad
\langle\langle A^*\rangle\rangle=A_0^*,
\label{sistema}
\end{eqnarray}
[with the inverse structure factor $S^{-1}(q,\eta)$ given by (\ref{structure})] 
and the spinodal of the N--N$_{\rm{B}}$ transition [$y(\kappa)$, 
given explicitly by Eq. (\ref{spin})] 
can be found by substituting (\ref{spin}) into 
(\ref{structure}) and solving (\ref{sistema}) for the variables 
$q$ and $\kappa$. Using the Fourier parameterization approach 
the Sm-Sm$_{\rm{B}}$ transition 
is located at the value of $\eta$ for which 
$s_{k1}\sim 0$, $\forall k$, and finally the N$_{\rm{B}}$--
Sm$_{\rm{B}}$ bifurcation must fulfill the condition $s_{1k}\sim 0$ $\forall k$.

Let us now prove that the four--phase point must occur.
In the neighborhood of this point, the leading order terms in the
order--parameter expansion of the free-energy difference
between the Sm$_{\rm{B}}$ and N phases from the Fourier parameterization 
approach [Eq. (\ref{cinco})] is
\begin{eqnarray}
\Delta\varphi=A(q,\eta;\kappa)s_{10}^2+B(\eta;\kappa)s_{01}^2,
\label{varphi}
\end{eqnarray}
where the coefficient $A(q,\eta;\kappa)$ is proportional to the inverse of the
structure factor $S^{-1}(q,\eta)$,
given by Eq. (\ref{structure}), with $\langle\langle A^*\rangle\rangle=A_0^*$,
while $B(\eta;\kappa)$ is
the coefficient of the quadratic term in the expansion of  
the free--energy difference between the N$_{\rm{B}}$ and N phases with 
respect to the Fourier amplitudes $h_i$, given
by Eq. (\ref{bifurca}).
Other terms in the expansion, depending on powers of $s_{11}$ and beyond, 
are of higher order in magnitude. Thus, at this order of approximation,
the smectic and orientational order parameters
are decoupled. Minimization of (\ref{varphi}) with respect to
$s_{10}$ and $s_{01}$ gives the set of coupled equations
\begin{eqnarray}
A(q^*,\eta;\kappa)=0,\quad B(\eta;\kappa)=0,
\label{tata}
\end{eqnarray}
with $q^*$ the value at the absolute minimum of $S^{-1}(q,\eta)$ with
respect to $q$. These equations should
be solved together for $\eta$ and $\kappa$ to find the (unique) point 
in the $\eta$--$\kappa$ plane where the N phase becomes unstable with
respect to Sm$_{\rm B}$ fluctuations. But note that
Eqn. (\ref{tata}), as pointed out before,
are also the equations, Eqns. (\ref{spin})-(\ref{sistema})
[obtained from the decoupling approximation], whose
simultaneous solution corresponds to the point where
the N-Sm and N-N$_{\rm{B}}$ spinodal curves meet. This proves that
this point is indeed a four-phase point.

\section{Expressions for the coefficients $A_{2n}^*$}
\label{AA}
In this section we provide the analytic expressions for the coefficients 
$A^*_{2n}$ corresponding to all the 
particle transverse sections studied.
\begin{eqnarray}
\text{R}:\quad A_{2n}^*&=&\frac{1}{\pi}\left\{\left(\kappa+\kappa^{-1}+2\right)\
\delta_{n0}
\right.\nonumber\\&-&\left.\frac{1}{2}\left[\kappa+\kappa^{-1}+2(-1)^n\right]
\frac{(1-\delta_{n0})}{4n^2-1}\right\},\\ \nonumber\\
\text{SDR}: \quad A_{2n}^*&=&\left(\kappa-\frac{1}{2}+\frac{\pi}{8}\right)^{-1}
\left\{\frac{1}{2}\left(\kappa+\frac{\pi}{8}+\frac{2\kappa^2}{\pi}
\right)\delta_{n0}\right.\nonumber\\
&-&\left.\frac{1}{2\pi}\left(\kappa+\frac{(-1)^n-1}{2}\right)^2
\frac{(1-\delta_{n0})}{4n^2-1}\right\},
\\ \nonumber \\
\text{DR}: \quad A^*_{2n}&=&
\left[1+\frac{(\kappa-1)^2}{\displaystyle{\pi\left(\kappa-1+\frac{\pi}{4}
\right)}}\right]\delta_{n0}\nonumber\\
&-&\frac{(\kappa-1)^2(1-\delta_{n0})}
{\displaystyle{2\pi\left(\kappa-1+\frac{\pi}{4}\right)(4n^2-1)}},\\
\nonumber \\
\text{D:} \quad A^*_{2n}&=&\frac{\left(\kappa+\kappa^{-1}\right)}{\pi}
\left[2\delta_{n0}-\cos(2n\gamma)^2\frac{(1-\delta_{n0})}{4n^2-1}\right],
\nonumber\\
\end{eqnarray}
with $\gamma=\arctan\kappa^{-1}$. Finally, the expression for the  E geometry  
can be calculated only numerically from 
\begin{eqnarray}    
E: \quad
A^*_{2n}&=&\frac{(1+\delta_{n0})}{2\pi}\int_0^{\pi}d\phi\cos(2n\phi)A^*(\phi),
\nonumber\\ A^*(\phi)&=&\frac{\left[s_1(\phi)+s_2(\phi)\right]}{\pi\kappa}
E[\tau(\phi)],
\end{eqnarray}
with $E(x)=\int_0^{\pi/2}dt \sqrt{1-x\sin^2t}$ 
the complete elliptic integral of the second kind and where there were defined 
\begin{eqnarray}
s_1(\phi)&=&
(\kappa^2-1)|\sin\phi|,\quad s_2(\phi)=\sqrt{4\kappa^2+s_1(\phi)^2},
\nonumber\\
\tau(\phi)&=&\frac{4s_1(\phi)s_2(\phi)}{\left[s_1(\phi)+s_2(\phi)\right]^2}.
\end{eqnarray} 
For $n=0$ we have $A^*_0=4\kappa [E(1-\kappa^{-2})]^2/\pi^2$.

\end{document}